\makeatletter
\def\eprint#1{\relax}
\def\@eprint#1{\relax}
\makeatother

\documentclass[10pt,twocolumn]{article}

\usepackage[letterpaper,margin=0.75in,columnsep=0.25in]{geometry}
\usepackage{cite}
\newcommand{\apj}{The Astrophysical Journal}

\usepackage{newtxtext,newtxmath}

\usepackage{graphicx}
\usepackage{floatrow}

\date{}

\makeatletter
\renewcommand{\fnum@figure}{\textbf{Figure \thefigure}}
\renewcommand{\fnum@table}{\textbf{Table \thetable}}
\makeatother

\usepackage{url}

\title{\bfseries \boldmath
The first detection of cosmic-ray excited H$_2$ in
interstellar space
}

\author{
    Shmuel Bialy$^{1\ast}$ 
    \and	
    Amit Chemke$^{1}$ 
    \and
    David A. Neufeld$^{2}$ 
    \and
    James Muzerolle Page$^{3}$ 
    \and
    Alexei V. Ivlev$^{4}$ 
    \and
    Sirio Belli$^{5}$ 
    \and
    Brandt A. L. Gaches$^{6}$ 
    \and
    Benjamin Godard$^{7,8}$ 
    \and
    Thomas G. Bisbas$^{9}$ 
    \and
    Paola Caselli$^4$ %
    \and
    Arshia M. Jacob$^{10,11}$ 
    \and
    Marco Padovani$^{12}$ 
    \and
    Christian Rab$^{13,4}$ 
    \and
    Kedron Silsbee$^{14}$ %
    \and 
    Troy A. Porter$^{15}$ 
    \and 
    \small$^{1}$Physics Department, Technion - Israel Institute of Technology, Haifa 3200003, Israel
    \\
    \small$^\ast$ Corresponding author. Email: sbialy@technion.ac.il
    \and
    \small$^{2}$William H. Miller III Department of Physics and Astronomy, The Johns Hopkins University, Baltimore, MD, USA
    \and
    \small$^{3}$Space Telescope Science Institute, Baltimore, MD, USA
    \and
    \small$^{4}$Max-Planck-Institut f\"ur extraterrestrische Physik, Giessenbachstrasse 1, 85748 Garching, Germany
    \and
    \small$^{5}$Dipartimento di Fisica e Astronomia, Università di Bologna, 40125 Bologna, Italy
    \and
    \small$^{6}$Faculty of Physics, University of Duisburg-Essen, Lotharstraße 1, 47057 Duisburg, Germany
    \and
    \small$^{7}$Observatoire de Paris, Université PSL, Sorbonne Université, LERMA, 75014 Paris, France
    \and
    \small$^{8}$Laboratoire de Physique de l’Ecole Normale Supérieure, ENS, Université PSL, CNRS, Sorbonne Université, Université de Paris, 75005 Paris, France
    \and
    \small$^{9}$Research Center for Astronomical Computing, Zhejiang Laboratory, Hangzhou, 311000, China
    \and 
    \small$^{10}$I. Physikalisches Institut, Universit\"at zu K\"oln, Z\"ulpicher Str. 77, D-50937 K\"oln, Germany
    \and 
    \small$^{11}$Max-Planck-Institut f\"ur Radioastronomie, Auf dem H\"ugel 69, 53121, Bonn, Germany
    \and
    \small$^{12}$INAF-Osservatorio Astrofisico di Arcetri, Largo E. Fermi 5, 50125 Firenze, Italy
    \and
    \small$^{13}$University Observatory, Faculty of Physics, Ludwig-Maximilians-Universit\"at M\"unchen, Scheinerstr. 1, 81679 Munich, Germany
    \and
    \small$^{14}$Physics Department, University of Texas at El Paso, El Paso 79968, USA
    \and 
    \small$^{15}$W. W. Hansen Experimental Physics Laboratory and Kavli Institute for Particle Astrophysics and Cosmology, Stanford University, \and
    \small Stanford, CA 94305, USA
}

\begin{document} 

\maketitle


{\boldmath \bfseries 
\noindent
Stars and planets form within cold, dark molecular clouds. In these dense regions, where starlight cannot penetrate, cosmic rays (CRs) are the dominant source of ionization---driving interstellar chemistry\cite{Dalgarno2006}, setting the gas temperature\cite{Goldsmith1969}, and enabling coupling to magnetic fields\cite{McKee2007}. Together, these effects regulate the collapse of clouds and the onset of star formation. Despite this importance, the cosmic-ray ionization rate, $\zeta$, has never been measured directly. Instead, this fundamental parameter has been loosely inferred from indirect chemical tracers and uncertain assumptions, leading to published values that span nearly two orders of magnitude and limiting our understanding of star formation physics.
Here, we report the first direct detection of CR-excited vibrational H$_2$ emission, using \textit{James Webb Space Telescope} (JWST) observations of the starless core Barnard 68 (B68). The observed emission pattern matches theoretical predictions for CR excitation precisely, confirming a decades-old theoretical proposal long considered observationally inaccessible. This result enables direct measurement of $\zeta$, effectively turning molecular clouds into natural, light-year-sized, cosmic-ray detectors. It opens a transformative observational window into the origin, propagation, and role of cosmic rays in star formation and galaxy evolution.
}

\begin{figure*}
\raggedright 
\floatbox[{\capbeside\thisfloatsetup{capbesideposition={right,top},capbesidewidth=5cm}}]{figure}[\FBwidth]
{\caption{
\textbf{H$_2$ Excitation.}  
The H$_2$ energy level diagram (left) shows rotational states in the $v=0$ and $v=1$ vibrational manifolds, split into para-H$_2$ (even $J$) and ortho-H$_2$ (odd $J$). Cosmic rays (CRs) preferentially excite $v=1$, $J=0$ and $J=2$, enhancing four key infrared lines: 1--0 S(0), Q(2), O(2), and O(4). This produces a characteristic spectrum (upper-right) with a peak at 2.63~$\mu$m (O(2)). In contrast, ultraviolet (UV) excitation (lower-right) yields a more uniform distribution across even and odd $J$.  
Spectra show relative intensities (arbitrary scale) to highlight excitation patterns.
}
\label{fig0}}
{\includegraphics[width=0.65\textwidth]{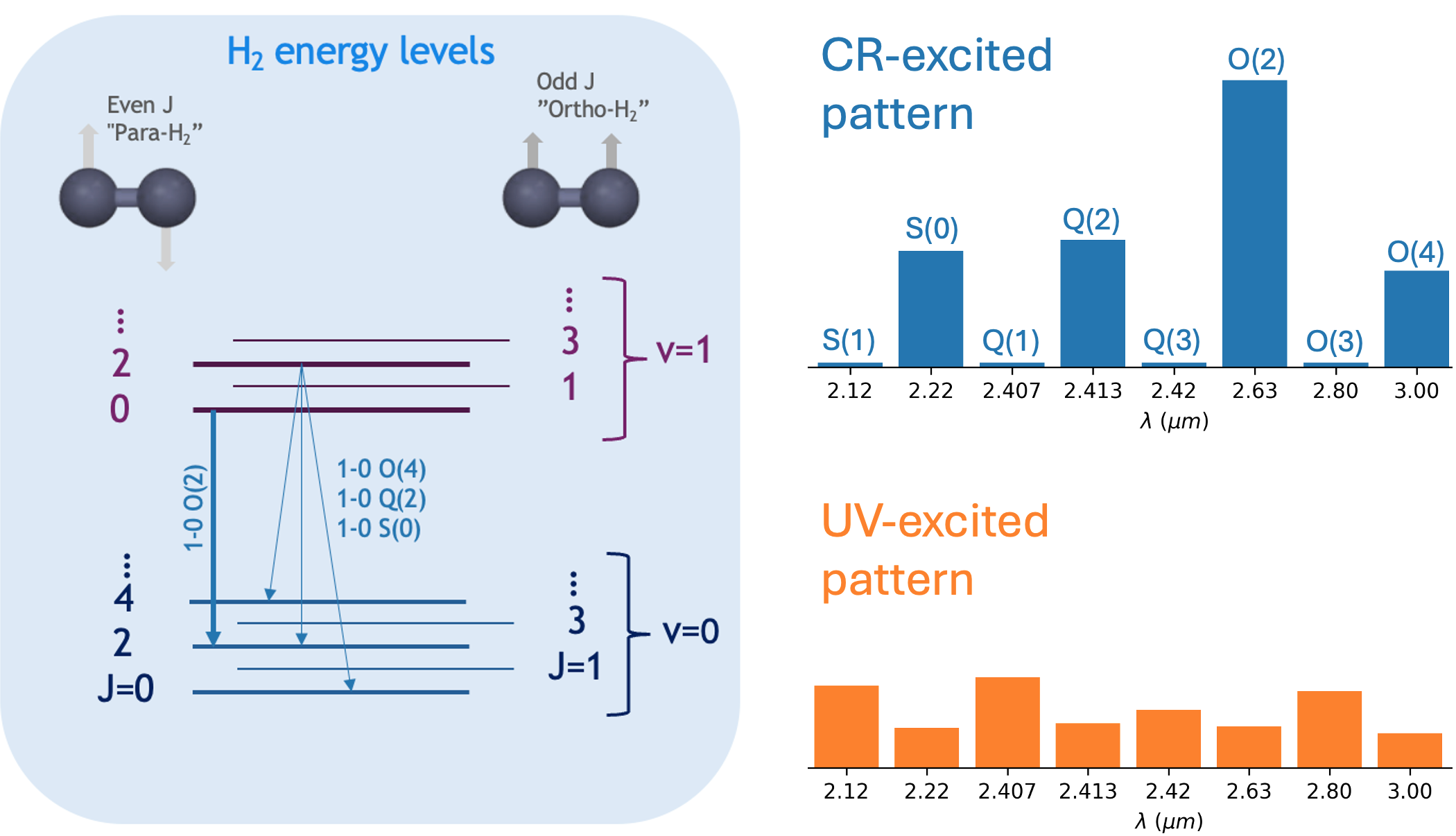}}
\end{figure*}

All previous measurements of $\zeta$\footnote{We define $\zeta$ (s$^{-1}$) to be the total ionization rate of H$_2$, including ionization by the primary CRs and additional ionization by the secondary CR electrons.} have relied on indirect inference through emission and absorption spectroscopy of rare molecular ions, including H$_3^+$, OH$^+$, HCO$^+$, and H$_2$D$^+$ \cite{Caselli1998, McCall2003, Indriolo2012, Indriolo2015, Neufeld2017b, Bialy2019c, Bovino2020, luo2024}. 
These methods require chemical modeling to interpret the observations, introducing substantial uncertainties due to poorly constrained reaction rates, uncertain gas density structure, and limited background sightlines for absorption studies\cite{Gaches2019, Neufeld2024, Obolentseva2024, indriolo_in_prep}. 

The theoretical foundation for an alternative approach began with Gredel et al.\cite{Gredel1995}, who showed that energetic electrons could excite H$_2$ molecules into ro-vibrational states. 
Building on this work, Bialy\cite{Bialy2020} first proposed that cosmic ray secondary electrons would similarly excite H$_2$, producing characteristic infrared emission lines that could be observed to directly measure $\zeta$ without chemical modeling. 
Padovani et al.\cite{padovani2022} further developed this theoretical framework with detailed quantum mechanical calculations of electron-H$_2$ cross sections.
This direct H$_2$ method offers several major advantages over traditional techniques: it probes H$_2$, the dominant molecular species, eliminating the need for chemical modeling; as an emission-based technique, it does not rely on background continuum sources; and the emission is directly linked to $\zeta$, with line intensities scaling linearly with the ionization rate because H$_2$ resides mostly in its ground state and undergoes rapid radiative decay following excitation.

\begin{figure*}[!t]
\raggedright 
\floatbox[{\capbeside\thisfloatsetup{capbesideposition={right,top},capbesidewidth=5cm}}]{figure}[\FBwidth]
{\caption{\textbf{Cosmic-ray-excited H$_2$ emission from Barnard 68 (B68).}  
\textbf{a,} Optical DSS2 red band image of B68. The white line indicates the JWST/NIRSpec slit location.   
\textbf{b,} Slit-integrated spectrum demonstrates two key signatures of cosmic-ray excitation: preferential enhancement of even-$J$ transitions 1-0 O(2), 1-0 Q(2), and 1-0 O(4) (blue bands) over adjacent odd-$J$ lines, and the predicted intensity hierarchy with O(2) as the dominant transition.
\textbf{c,} The best-fit excitation model ($\chi^2_{\rm reduced} = 1.5$) yields a cosmic-ray ionization rate of $\zeta = (1.7 \pm 0.1) \times 10^{-16}\ \mathrm{s}^{-1}$ and UV radiation intensity $\chi_{\rm UV}=1.89 \pm 0.04$, reproducing both the absolute line strengths and diagnostic ratios. The orange bars show the contribution from UV excitation alone, while CR excitation (blue bars) is required to explain the elevated 1-0 O(2), 1-0 Q(2), and 1-0 O(4) line intensities.}
\label{fig1}}
{\includegraphics[width=0.65\textwidth]{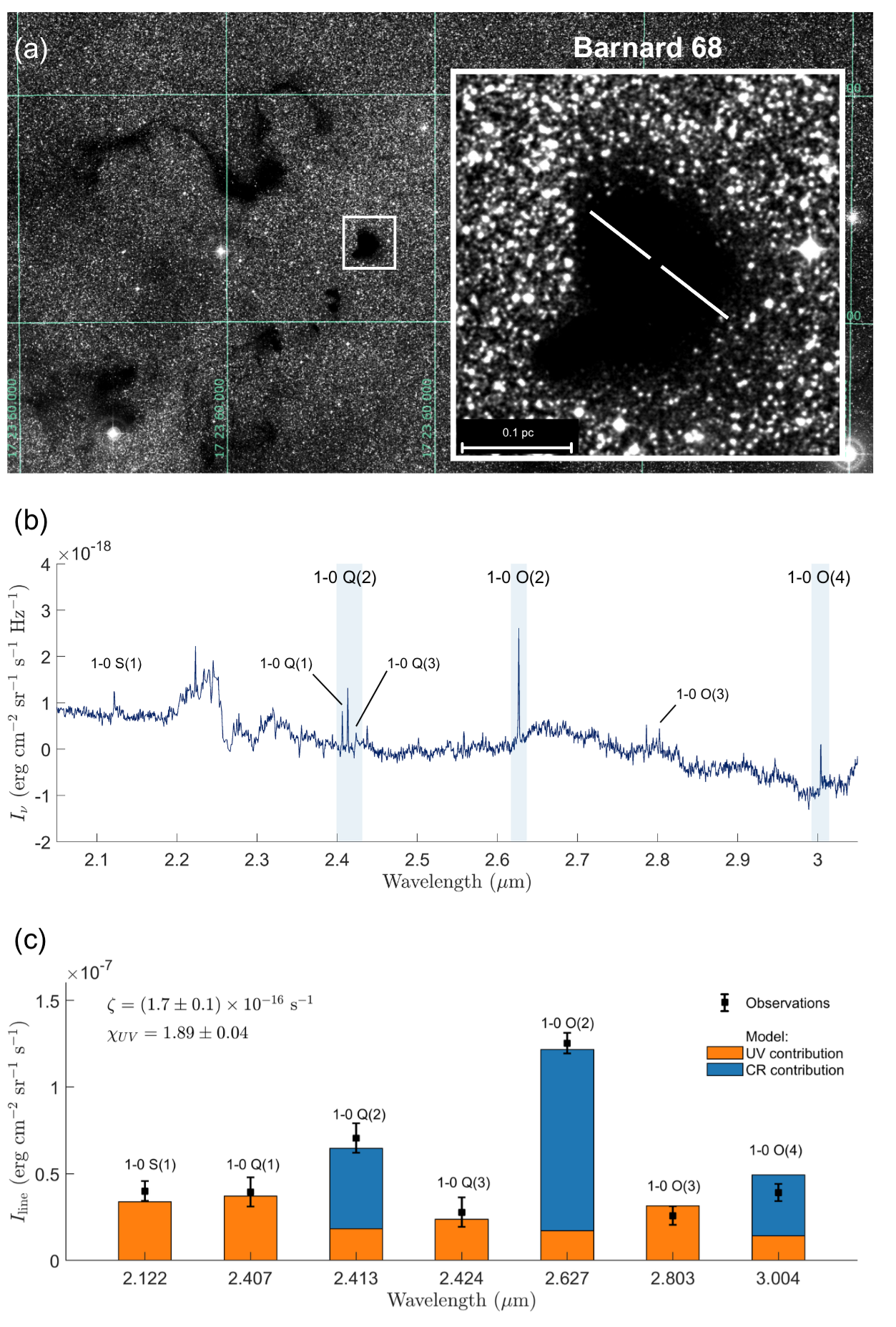}}
\end{figure*}

However, reliable application of this method requires
distinguishing CR excitation from competing processes. 
Given the large vibrational energy gap between the ground ($v=0$) and first excited ($v=1$) vibrational states ($\Delta E/k_B \approx 6000$~K \cite{Roueff2019}), thermal excitation in cold clouds like B68 (where $T \approx 10$~K \cite{hotzel2002, lada2003}) is negligible. 
This leaves two relevant non-thermal mechanisms: UV and CR excitation.
Crucially, UV and CR excitation produce different emission patterns, reflecting their distinct physical origins.

Primary CR protons penetrate deep into the cloud interior, generating secondary electrons that excite rotational and vibrational levels of H$_2$ via direct electron impact.
The low temperatures in these shielded regions maintain H$_2$ predominantly in its ground $v=0$, $J=0$ state (para-H$_2$).
Following quantum selection rules, CR secondary electrons can only excite even $J$ states, with excitation primarily to the $v=1$, $J=0$ and $v=1$, $J=2$ levels \cite{Gredel1995}.
Radiative decay from these levels produces emission in four specific near-infrared lines: 1–0 S(0), 1-0 Q(2), 1-0 O(2), and 1-0 O(4) at 2.22, 2.41, 2.63, and 3.0 $\mu$m, respectively, with the 1-0 O(2) line being $2.3\text{–}3$ times stronger than the others.
This yields a distinctive CR signature: strong, concentrated emission in these four even-$J$ transitions (Fig.~\ref{fig0}).

UV excitation operates through a fundamentally different mechanism and spatial regime.
It proceeds via fluorescence—a two-step process in which H$_2$ molecules are first excited to electronic states and then radiatively cascade to vibrationally excited levels of the ground electronic state.
UV fluorescence occurs in the outer cloud envelopes, where UV photons can penetrate before being absorbed by dust.
In these warmer outer regions, thermal energy populates both para-H$_2$ ($v=0$, $J=0$) and ortho-H$_2$ ($v=0$, $J=1$) initial states. 
Fluorescence from these mixed populations can excite both even and odd $J$ levels in the $v=1$ manifold, producing a broad emission spectrum with nearly uniform intensities across even- and odd-$J$ transitions—a diagnostic signature that clearly distinguishes UV from CR excitation (Fig.~\ref{fig0}).

 \subsection*{Detecting CR-Excited H$_2$ with the James Webb Space Telescope}

While long theorized, the faint nature of these emission lines has so far prevented direct observation of CR-excited H$_2$, with ground-based attempts failing to detect these emissions \cite{Bialy2022}. The launch of JWST and its extraordinary sensitivity in the near-infrared has now transformed this observational landscape. Leveraging this unprecedented capability, we used the Near Infrared Spectrograph (NIRSpec) to observe the H$_2$ emission spectrum of B68, a well-studied, nearby starless core (Fig.~\ref{fig1}a).

We selected B68 as our target due to its optimal physical characteristics: a high molecular hydrogen column density ($N({\rm H_2}) = (1-4) \times 10^{22}$~cm$^{-2}$\cite{Nielbock2012, Roy2014}), low gas temperature ($T \approx 10$~K\cite{hotzel2002, lada2003}), and weak ambient UV irradiation\cite{Roy2014, Lippok2016}. These properties enhance the relative contribution of CR-excited emission, compared to UV and thermal excitation processes.
Our observational strategy utilized background-subtracted spectra to minimize instrumental noise and remove contamination from zodiacal and intervening gas emission. We further integrated the spectra along the slit to optimize the signal-to-noise ratio.

Our observations reveal the distinct spectral signature of CR excitation for the first time (Fig.~\ref{fig1}b). Three of the four\footnote{The 1-0 S(0) line intensity could not be reliably determined because, for a significant fraction of positions along the slit, its wavelength fell within the gap between the NIRSpec detector arrays.} CR-dominated transitions are unambiguously detected: 1–0 O(2), 1–0 Q(2), and 1–0 O(4), and are significantly enhanced relative to neighboring odd-$J$ transitions. Among these three transitions, the 1–0 O(2) line shows a $\sim2.5$ times enhancement over the Q(2) and O(4) lines, closely matching theoretical predictions of a 2.3–3 $\times$ intensity ratio.


\subsection*{Derivation of the CR ionization rate}

To quantify the CR ionization rate, we applied the theoretical framework developed by Bialy \cite{Bialy2020}, incorporating updated quantum mechanical calculations of electron-impact cross-sections from Padovani et al. \cite{padovani2022} to the observed line intensities (Fig.~\ref{fig1}c).

The model contains only two free parameters, the CR ionization rate $\zeta$ and the normalized UV field energy density, $\chi_{\mathrm{UV}} \equiv u_{\mathrm{UV}} / u_{\mathrm{UV,0}}$, where $u_{\mathrm{UV,0}}=0.056$~eV~cm$^{-3}$ is the standard interstellar radiation energy density in the solar neighborhood \cite{Draine1978, Bialy2020b}.
Each excitation mechanism produces a fixed spectral pattern—specific line intensity ratios determined by microphysics alone, independent of the values of $\zeta$ and $\chi_{\rm UV}$. The $\zeta$ and $\chi_{\rm UV}$ parameters simply scale the overall spectrum brightness up or down (Eqs. \ref{eq: CRs}-\ref{eq: I_UV}), preserving the characteristic pattern of each excitation mechanism (Fig.~\ref{fig0} - right panels; see also Methods section).

\begin{figure}
\includegraphics[width=\textwidth]{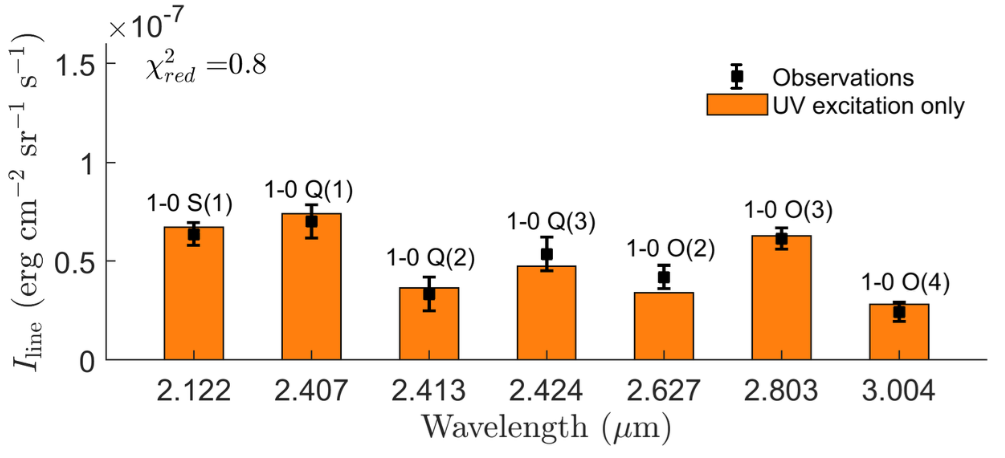}
\caption{
\textbf{Background region confirms UV excitation model.}
JWST/NIRSpec slit-integrated H$_2$ line intensities in our background observation, a diffuse, low-column-density region (black points), compared to the best-fit UV excitation model (orange bars). The excellent agreement demonstrates that UV excitation alone accounts for the observed emission, with an emission pattern that matches theoretical predictions ($\chi^2_{\rm red} = 0.8$). This independent validation strengthens our theoretical framework and supports the identification of CR excitation in B68.
}
\label{fig2}
\end{figure}

Applying our CR+UV model, we derive a CR ionization rate of 
\begin{equation}
    \zeta = (1.7 \pm 0.1) \times 10^{-16} \,\mathrm{s}^{-1}
\end{equation}
for B68, along with a far-UV field strength of $\chi_{\mathrm{UV}} = 1.89 \pm 0.04$. Both values represent spatial averages across the slit coverage (Fig.~\ref{fig1}c). To assess the statistical significance of this result, we have considered a competing model where the H$_2$ is excited purely by UV. The model that includes CR excitation provides a significantly better fit with $\chi^2_{\mathrm{reduced}} = 1.5$ vs 55.8 (see Fig.~\ref{fig3}), with an F-test probability $P < 10^{-4}$ and an Akaike information criterion $\Delta AIC = 325$, conclusively rejecting the pure UV hypothesis. 

To further validate our approach, we observed a nearby ``background'' region with significantly lower column density ($\approx 10 \times$ lower than B68). Since in the optically thin limit CR excitation scales proportionally with column density\cite{Bialy2020}), this region should exhibit negligible CR excitation, allowing us to test our UV model in isolation. With only one free parameter ($\chi_{\mathrm{UV}}$), the pure-UV model fits this background observation remarkably well (Fig.~\ref{fig2}), validating our theoretical framework for UV excitation.

Our derived CR ionization rate provides valuable context when compared to the current benchmark set by H$_3^+$ absorption measurements. A recent analysis across 16 sight lines in the local interstellar medium reports values ranging from $\zeta = 2.5 \times 10^{-17}$ to $1.2 \times 10^{-16}$ s$^{-1}$ in clouds with the total hydrogen column densities between $1.6 \times 10^{21}$ and $4.9 \times 10^{21}$ cm$^{-2}$ \cite{Obolentseva2024, indriolo_in_prep}. Our value lies near the upper end of this range. 
This suggests that B68 may be exposed to an enhanced local CR flux due to nearby CR acceleration sites. 

While H$_3^+$ observations provide robust estimates, they require knowledge of uncertain volume densities and cloud structure and necessitate hot background stars for absorption line measurements. Our H$_2$ emission approach addresses these observational constraints by directly probing the dominant molecular component without requiring background sources or uncertain density estimates.
Rather than replacing H$_3^+$ observations, the two methods provide complementary diagnostic capabilities across different density regimes: H$_3^+$ absorption traces diffuse molecular gas, while H$_2$ emission accesses the dense regions ($N({\rm H_2}) \gtrsim 10^{22}$ cm$^{-2}$). This complementarity opens new possibilities for systematic studies. As the H$_2$ method expands to additional molecular clouds across the Galaxy, the synergy of the two methods will enable comprehensive mapping of CR ionization rates as a function of gas density and environment. This will provide powerful new constraints on CR propagation and transport mechanisms throughout the interstellar medium.

\subsection*{Conclusions}

The CR ionization rate is a fundamental parameter governing the thermal, chemical, and dynamical evolution of astrophysical gas across a broad range of environments. In cold, dense cores, where stars begin to form, CRs control the ionization fraction, coupling between gas and magnetic fields, and heating, thereby regulating core collapse and fragmentation. On galactic scales, particularly in low-metallicity systems, CRs are expected to dominate ISM heating and drive the conversion of atomic to molecular gas, shaping star formation and the evolution of dwarf galaxies and early-universe systems \cite{Bialy2015a, Bialy2019, Sternberg2021}. A direct and robust measurement of the CR ionization rate thus provides a critical observational anchor for models linking small-scale cloud physics to galaxy-scale star formation regulation.

Our results establish a new observational framework in which molecular clouds act as natural CR detectors, revealing CR ionization through H$_2$ excitation and emission. This method circumvents key limitations of traditional approaches to measure $\zeta$ by directly probing the bulk molecular gas, without relying on rare background sources or uncertain assumptions on density structure and chemical reaction rates. As JWST observations expand to additional regions of the Galaxy, this technique will enable systematic mapping of the CR ionization rate across diverse environments, placing strong constraints on CR origins, propagation, and feedback. 

Much like particle detectors on Earth, molecular clouds now serve as natural, gigantic detectors floating in space and spanning light-years—opening a powerful new avenue to trace CRs, from local to galactic scales.


\clearpage 

%

%
%
%
%
%
%


\section*{Acknowledgments}
This work is based on observations made with the NASA/ESA/CSA James Webb Space Telescope obtained through observing program GO-5064. We acknowledge the dedicated efforts of the Space Telescope Science Institute staff in scheduling and executing these observations. The James Webb Space Telescope is operated by the Space Telescope Science Institute under NASA contract NAS5-03127.

We thank Joao Alves and Charlie Lada for providing us with column density data for the B68 cloud. SB thanks Amiel Sternberg for helpful discussions. 
SB thanks Oren Slone for an interesting discussion on Dark Matter and its possible interactions with H$_2$ which sparked the idea behind the ``direct H$_2$ method''.

\paragraph*{Funding:}
SB acknowledges financial support from the Israeli Science Foundation grant 2071540, and the German Israeli Science Foundation Grant 1568, and the Technion.
CHR acknowledge the support of the Deutsche Forschungsgemeinschaft (DFG, German Research Foundation) Research Unit ``Transition discs'' - 325594231. CHR, AVI and PC are grateful for support from the Max Planck Society. BALG is supported by the German Deutsche Forschungsgemeinschaft, DFG, in the form of an Emmy Noether Research Group - Project-ID 542802847. SBe is supported by the European Research Council (StG 101076080). TGB acknowledges support from the Leading Innovation and
Entrepreneurship Team of Zhejiang Province of China (Grant No. 2023R01008).
DAN, KS, and TP gratefully acknowledge support from USRA grant JWST-GO-5604. BG would like to acknowledge the support from the Programme National “Physique et Chimie du Milieu Interstellaire” (PCMI) of CNRS/INSU with INC/INP co-funded by CEA and CNES.
TAP acknowledges partial funding via NASA grants 80NSSC22K0477, 80NSSC22K0718, and 80NSSC23K0169, and USRA grant JWST-GO-5604.
\paragraph*{Competing interests:}
There are no competing interests to declare.
\paragraph*{Data and materials availability:}
Raw and processed data are available through the MAST archive under program ID GO 5064. Reduced spectra, line measurements, and derived parameters are provided in the supplementary materials.




\renewcommand{\thefigure}{M\arabic{figure}}
\renewcommand{\thetable}{M\arabic{table}}
\renewcommand{\theequation}{M\arabic{equation}}
\setcounter{figure}{0}
\setcounter{table}{0}
\setcounter{equation}{0}







\section*{Methods}

\subsection*{Observational Strategy}

Observations were conducted using JWST/NIRSpec in Cycle 3 (GO 5064). The Micro-Shutter Assembly (MSA) was configured as a long slit by opening all shutters along a column, creating a 3.4 arcmin aperture comparable to B68's angular extent. We adopted the G235H grating with F170LP filter, providing spectral resolution $R \approx 2700$ that enables clear separation of diagnostic H$_2$ lines, including the closely spaced 1-0 Q(1) and 1-0 Q(2) transitions critical for distinguishing UV- and CR-induced excitation mechanisms.

Target observations employed five dithered positions with the NRSIRS2 readout pattern (4 integrations of 20 groups each), yielding 8.2 hours total integration time. The dithering strategy stepped 5 shutters in both spatial directions to minimize stellar contamination and mitigate bad pixel effects.

Our observational strategy employed background subtraction, with the background region positioned 30 arcmin from B68's center, sampling material with a column density approximately 10 times lower than that of B68. The small column density ensures negligible CR excitation in the background region, as CR excitation scales with $N$ in the optically thin limit (Eq.~\ref{eq: CRs}). 
The background observations used three dithered positions with single integrations of 20 groups each (1.2 hours total).

The background region serves two roles.
First, it removes line-of-sight contamination from zodiacal light and from intervening interstellar gas excited by the ambient interstellar UV radiation field.
The background-subtracted spectrum thus isolates the H$_2$ excitation (from both UV and CRs) that occurs specifically in B68 (and its envelope). 
Hence in our main analysis (Fig.~\ref{fig1}) we utilize the background-subtracted spectra.
Second, since the background region consists of interstellar gas that is excited by the ambient UV radiation field only, it provides excellent observational data for observationally validating our UV excitation model (Eq.~\ref{eq: I_UV}).
We perform this validation in our main analysis (Fig.~\ref{fig3}).

\subsection*{Data Reduction}

All exposures were reprocessed starting from raw "uncal" files using JWST pipeline version 1.16.1 and Calibration Reference Data System context 1303. Default parameters were employed for all processing steps except for outlier detection in the stage 3 pipeline, where "exptime" weighting yielded a significant decrease in unmasked outliers compared to standard settings. 

We modified the MSA metadata to enable the pipeline to combine and extract spectra for each open shutter of the long slit across all dithered positions for both target and background observations. This approach generated approximately 340 individual 1D spectra per observation.

Integrated spectra for both target and background observations were generated by combining all single-shutter spectra, manually excluding those showing clear contamination from background galaxies or foreground stars.

\subsection*{Quantitative Model: Extracting $\zeta$ from H$_2$ Emission}

To quantitatively separate CR and UV contributions and determine the CR ionization rate ($\zeta$), we developed a framework based on \cite{Bialy2020}. The total line intensity for any ro-vibrational transition combines both contributions:
\begin{equation}
\label{eq: Itot}
    I_{\rm ul} = I_{\rm ul}^{\rm (CR)} + I_{\rm ul}^{\rm (UV)}
\end{equation}
where the subscripts $u$ and $l$ indicate the upper and lower energy states, as determined by their vibrational and rotational quantum numbers $v$, $J$.

\textbf{CR Component:} CR excitation generates emission intensity given by\cite{Bialy2020}:
\begin{equation}
\label{eq: CRs}
    I_{\rm ul}^{\rm (CR)} = \frac{1}{4 \pi} b_{\rm u}\zeta \ \alpha_{\rm ul} E_{\rm ul}\ N({\rm H_2}) g
\end{equation}
where $\zeta$ is the CR ionization rate and $b_{\rm u} \equiv \zeta_{\rm ex, u}/\zeta$ represents the entry probabilities into level $u$  determined from recently calculated level-by-level inelastic electron-impact cross sections \cite{padovani2022}. The branching ratios $\alpha_{\rm ul}$ for radiative decay are determined by ratios of Einstein A coefficients, $E_{\rm ul}$ is the transition energy, and $N({\rm H_2})$ is the H$_2$ column density.

The factor $g=(1-\mathrm{e}^{-\tau})/\tau$ accounts for dust attenuation, where $\tau=0.9 N({\rm H_2})/(10^{22} \ {\rm cm^{-2}})$ is the dust opacity in the 2-3 $\mu$m range \cite{Draine2011, Bialy2020}. We derive $N({\rm H_2})$ from a high-resolution column density map using the NICER extinction method \cite{Lombardi2001, Alves2001}, assuming $A_V/N_H=5.3\times10^{-22}$ mag cm$^2$\cite{Draine2011}, where $N_H$ is the hydrogen nucleus number density, and $N_H=2N({\rm H_2})$. Our neglect of atomic hydrogen is justified because the H/H$_2$ ratio becomes negligibly small in high column density regions \cite{Sternberg2014, Bialy2016a}.
Using this $N({\rm H_2})$ map, we calculate $g$ and $N({\rm H_2})$ at each position along the slit and compute the slit-averaged value $\langle gN({\rm H_2}) \rangle=6.5 \times 10^{21}$ cm$^{-2}$ for B68. This value is used to derive the slit-averaged CR ionization rate $\zeta$ in Eq.~(\ref{eq: CRs}).

\textbf{UV Component:} 
UV excitation operates through a two-step fluorescence process. First, H$_2$ electronic Lyman and Werner states are excited, followed by radiative cascades that populate vibrational levels in the ground electronic state. These excited vibrational states then emit the observed infrared lines. For high-density conditions like those in B68, analytic models predict that the intensity of UV-excited H$_2$ emission scales linearly with the UV radiation field strength\cite{Sternberg1988, black1987a}. Following Ref.~\cite{Bialy2020} we adopt the expression:
\begin{equation}
\label{eq: I_UV}
    I_{\rm ul}^{\rm (UV)} =  9.6 \times 10^{-7} \chi_{\rm UV} f_{\rm ul} \ {\rm erg \ cm^{-2} \ s^{-1} \ sr^{-1}} \ , 
\end{equation}
where the prefactor $9.6 \times 10^{-7} \chi_{\rm UV}$ ${\rm erg \ cm^{-2} \ s^{-1} \ sr^{-1}}$ is the total intensity in UV-excited H$_2$ emission lines, and the relative line intensities, $f_{\rm ul}$ (such that $\Sigma_{ul} f_{\rm ul}=1$),  are mostly independent of the environmental conditions and are determined primarily by the Einstein A coefficients for radiative decay\cite{Sternberg1988}.
We have further verified Eq.~(\ref{eq: I_UV}) against detailed numerical results of the photodissociation region (PDR) model of Meudon\cite{LePetit2006}.

The entry probabilities $b_{\rm ul}$ and the $f_{\rm ul}$ values are presented in Table \ref{tab:atomicInfo}.

\subsection*{Spectral Fitting and Parameter Determination}

We identify H$_2$ emission lines as statistically significant peaks at wavelengths coinciding with known H$_2$ transition\cite{Roueff2019}. For each detected line, we fitted and subtracted a local linear baseline to isolate the emission signal. Integrated line intensities were computed by summing the flux across the full line width in the continuum-subtracted spectrum. Measurement uncertainties were estimated from the standard deviation of the continuum noise within $\Delta \lambda = 0.01\,\mu m$  range on both sides of each line, scaled by the square root of the number of integration pixels in the emission line, and added in quadrature to the instrumental noise provided by the JWST/NIRSpec pipeline. Gaussian profile fitting was performed as validation, yielding results consistent with direct flux summation.

Our fitting procedure employs $\chi^2$ minimization to simultaneously model all detected H$_2$ line intensities using the theoretical framework described above. The optimization determines optimal values for $\zeta$ and $\chi_{\rm UV}$ while propagating uncertainties through the full parameter space to establish robust confidence intervals.

\begin{figure}
\includegraphics[width=\textwidth]{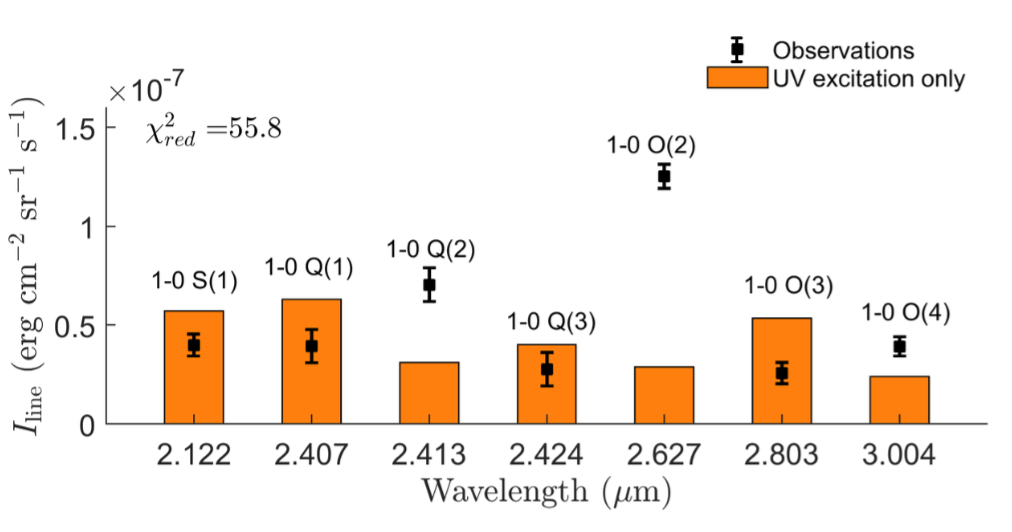}
\caption{\textbf{Inadequacy of the pure UV model for explaining H$_2$ excitation in B68.} H$_2$ line intensities observed toward B68 (data points with error bars) compared to predictions from a pure UV excitation model (orange bars). The model fails to adequately reproduce the observations, particularly underestimating the Q(2), O(2), and O(4) line intensities. The pure UV model yields $\chi^2_{\rm reduced} = 55.8$ with an F-test probability $P < 10^{-4}$, conclusively rejecting the pure UV hypothesis, demonstrating the necessity of CR excitation to explain the observations (see Fig.~\ref{fig1}c).}
\label{fig3}   
\end{figure}

We further evaluated the model adequacy by comparing our two-component CR+UV excitation framework against simpler single-component alternatives using standard statistical criteria: $\chi^2$ analysis, Fisher F-tests, and Akaike Information Criterion.
The pure UV model proves inadequate for the B68 observations. This model yields $\chi^2_{\rm reduced} = 55.8$ (Fig.~\ref{fig3}) compared to $\chi^2_{\rm reduced} = 1.5$ for the UV+CR model (Fig.~\ref{fig1}), corresponding to an F-test probability $P < 10^{-4}$ and $\Delta AIC = 325$, conclusively rejecting the pure UV hypothesis.


\begin{table*}[t]
\centering
    \caption{Spectroscopic parameters for H$_2$ $v=1 \rightarrow 0$ vibrational transitions used in our  model. The entry probabilities $b$ are based on quantum mechanical calculations of electron-impact cross sections \cite{Padovani2022a} at $T=10$~K, updating previous estimates from \cite{Gredel1995}. The UV excitation fractions $f$ are from \cite{Sternberg1988}, verified against comprehensive PDR model computations \cite{lepetit2006a}. For the first four transitions (CR-excited lines), $b$ values are insensitive to temperature. For the remaining odd-$J'$ transitions, $b$ values are temperature-sensitive, but cosmic ray excitation remains negligible as these lines are strongly dominated by UV excitation. The 1-0 S(0) transition was not included in the model due to unavailable observational data for this transition.}
        \label{tab:atomicInfo}
\begin{tabular}{|c|c|c|c|c|c|c|c|c|c|c|}
    \hline
    \textbf{Transition} & $\boldsymbol{v_u}$& $\boldsymbol{J_u}$& $\boldsymbol{v_l}$& $\boldsymbol{J_l}$& $\boldsymbol{\lambda\,(\mu\mathrm{m})}$ & $\boldsymbol{E\,(\mathrm{eV})}$ & $\boldsymbol{A\,(\mathrm{s}^{-1})}$ & $\boldsymbol{\alpha_{ul}}$& $\boldsymbol{b_u}$& $\boldsymbol{f_{ul}}$\\
    \hline
    (1-0) O(2) & 1 & 0 & 0 & 2 & 2.63 & 0.47 & $8.56 \times 10^{-7}$ & 1.00 & 1.57 & $9.4 \times 10^{-3}$ \\
    (1-0) Q(2) & 1 & 2 & 0 & 2 & 2.41 & 0.51 & $3.04 \times 10^{-7}$ & 0.36 & 1.78 & $1.01 \times 10^{-2}$ \\
    (1-0) S(0) & 1 & 2 & 0 & 0 & 2.22 & 0.56 & $2.53 \times 10^{-7}$  & 0.30 & 1.78 & $9.1\times10^{-3}$ \\
    (1-0) O(4) & 1 & 2 & 0 & 4 & 3.00 & 0.41 & $2.91 \times 10^{-7}$ & 0.34 & 1.78 & $7.8 \times 10^{-3}$ \\
    (1-0) Q(1) & 1 & 1 & 0 & 1 & 2.41 & 0.52 & $4.30 \times 10^{-7}$ & 0.50 & $8.2 \times 10^{-7}$ & $2.05 \times 10^{-2}$ \\
    (1-0) S(1) & 1 & 3 & 0 & 1 & 2.12 & 0.58 & $3.48 \times 10^{-7}$ & 0.42 & $3.7 \times 10^{-7}$ & $1.86 \times 10^{-2}$ \\
    (1-0) O(3) & 1 & 1 & 0 & 3 & 2.80 & 0.44 & $4.24 \times 10^{-7}$ & 0.50 & $8.2 \times 10^{-7}$ & $1.74 \times 10^{-2}$ \\
    (1-0) Q(3) & 1 & 3 & 0 & 3 & 2.42 & 0.51 & $2.79 \times 10^{-7}$ & 0.33 & $3.7 \times 10^{-7}$ & $1.31 \times 10^{-2}$ \\
    \hline
    \end{tabular}
\end{table*}

\clearpage 






\begin{thebibliography}{10}
\expandafter\ifx\csname url\endcsname\relax
  \def\url#1{\texttt{#1}}\fi
\expandafter\ifx\csname urlprefix\endcsname\relax\def\urlprefix{URL }\fi
\providecommand{\bibinfo}[2]{#2}
\providecommand{\eprint}[2][]{\url{#2}}

\bibitem{Dalgarno2006}
\bibinfo{author}{Dalgarno, A.}
\newblock \emph{\bibinfo{journal}{Proceedings of the National Academy of Sciences of the United States of America}} \textbf{\bibinfo{volume}{103}}, \bibinfo{pages}{12269--73} (\bibinfo{year}{2006}).

\bibitem{Goldsmith1969}
\bibinfo{author}{Goldsmith, D.~W.}, \bibinfo{author}{Habing, H.~J.} \& \bibinfo{author}{Field, G.~B.}
\newblock \emph{\bibinfo{journal}{The Astrophysical Journal}} \textbf{\bibinfo{volume}{158}}, \bibinfo{pages}{173} (\bibinfo{year}{1969}).

\bibitem{McKee2007}
\bibinfo{author}{McKee, C.~F.} \& \bibinfo{author}{Ostriker, E.~C.}
\newblock \emph{\bibinfo{journal}{Annual Review of Astronomy and Astrophysics}} \textbf{\bibinfo{volume}{45}}, \bibinfo{pages}{565--687} (\bibinfo{year}{2007}).
\newblock \eprint{0707.3514}.

\bibitem{Caselli1998}
\bibinfo{author}{Caselli, P.}, \bibinfo{author}{Walmsley, C.~M.}, \bibinfo{author}{Terzieva, R.} \& \bibinfo{author}{Herbst, E.}
\newblock \emph{\bibinfo{journal}{The Astrophysical Journal}} \textbf{\bibinfo{volume}{499}}, \bibinfo{pages}{234--249} (\bibinfo{year}{1998}).

\bibitem{McCall2003}
\bibinfo{author}{McCall, B.~J.} \emph{et~al.}
\newblock \emph{\bibinfo{journal}{Nature}} \textbf{\bibinfo{volume}{422}}, \bibinfo{pages}{500--2} (\bibinfo{year}{2003}).

\bibitem{Indriolo2012}
\bibinfo{author}{Indriolo, N.} \& \bibinfo{author}{McCall, B.~J.}
\newblock \emph{\bibinfo{journal}{The Astrophysical Journal}} \textbf{\bibinfo{volume}{745}}, \bibinfo{pages}{91} (\bibinfo{year}{2012}).

\bibitem{Indriolo2015}
\bibinfo{author}{Indriolo, N.} \emph{et~al.}
\newblock \emph{\bibinfo{journal}{The Astrophysical Journal}} \textbf{\bibinfo{volume}{800}}, \bibinfo{pages}{40} (\bibinfo{year}{2015}).

\bibitem{Neufeld2017b}
\bibinfo{author}{Neufeld, D.~A.} \& \bibinfo{author}{Wolfire, M.~G.}
\newblock \emph{\bibinfo{journal}{The Astrophysical Journal}} \textbf{\bibinfo{volume}{845}}, \bibinfo{pages}{163} (\bibinfo{year}{2017}).
\newblock \eprint{1704.03877}.

\bibitem{Bialy2019c}
\bibinfo{author}{Bialy, S.}, \bibinfo{author}{Neufeld, D.}, \bibinfo{author}{Wolfire, M.}, \bibinfo{author}{Sternberg, A.} \& \bibinfo{author}{Burkhart, B.}
\newblock \emph{\bibinfo{journal}{The Astrophysical Journal}} \textbf{\bibinfo{volume}{885}}, \bibinfo{pages}{109} (\bibinfo{year}{2019}).
\newblock \eprint{1909.12305}.

\bibitem{Bovino2020}
\bibinfo{author}{Bovino, S.}, \bibinfo{author}{{Ferrada-Chamorro}, S.}, \bibinfo{author}{Lupi, A.}, \bibinfo{author}{Schleicher, D.~R.} \& \bibinfo{author}{Caselli, P.}
\newblock \emph{\bibinfo{journal}{Monthly Notices of the Royal Astronomical Society: Letters}} \textbf{\bibinfo{volume}{495}}, \bibinfo{pages}{L7--L11} (\bibinfo{year}{2020}).
\newblock \eprint{2003.05416}.

\bibitem{luo2024}
\bibinfo{author}{Luo, G.}, \bibinfo{author}{Bisbas, T.~G.}, \bibinfo{author}{Padovani, M.} \& \bibinfo{author}{Gaches, B. A.~L.}
\newblock \emph{\bibinfo{journal}{Astronomy \& Astrophysics}} \textbf{\bibinfo{volume}{690}}, \bibinfo{pages}{A293} (\bibinfo{year}{2024}).

\bibitem{Gaches2019}
\bibinfo{author}{Gaches, B. A.~L.}, \bibinfo{author}{Offner, S. S.~R.} \& \bibinfo{author}{Bisbas, T.~G.}
\newblock \emph{\bibinfo{journal}{The Astrophysical Journal}} \textbf{\bibinfo{volume}{878}}, \bibinfo{pages}{105} (\bibinfo{year}{2019}).
\newblock \eprint{1905.02232}.

\bibitem{Neufeld2024}
\bibinfo{author}{Neufeld, D.~A.} \emph{et~al.} (\bibinfo{year}{2024}).
\newblock \eprint{2408.11108}.

\bibitem{Obolentseva2024}
\bibinfo{author}{Obolentseva, M.} \emph{et~al.}
\newblock \emph{\bibinfo{journal}{The Astrophysical Journal}} \textbf{\bibinfo{volume}{973}}, \bibinfo{pages}{142} (\bibinfo{year}{2024}).

\bibitem{indriolo_in_prep}
\bibinfo{author}{Indriolo, N.} \emph{et~al.}
\newblock \emph{\bibinfo{journal}{in prep}} .

\bibitem{Gredel1995}
\bibinfo{author}{Gredel, R.} \& \bibinfo{author}{Dalgarno, A.}
\newblock \emph{\bibinfo{journal}{\apj}} \bibinfo{pages}{852} (\bibinfo{year}{1995}).

\bibitem{Bialy2020}
\bibinfo{author}{Bialy, S.}
\newblock \emph{\bibinfo{journal}{Nature Communications Physics}} \textbf{\bibinfo{volume}{3}}, \bibinfo{pages}{32} (\bibinfo{year}{2020}).

\bibitem{padovani2022}
\bibinfo{author}{Padovani, M.} \emph{et~al.}
\newblock \emph{\bibinfo{journal}{Astronomy \& Astrophysics}} \textbf{\bibinfo{volume}{658}}, \bibinfo{pages}{A189} (\bibinfo{year}{2022}).

\bibitem{Roueff2019}
\bibinfo{author}{Roueff, E.} \emph{et~al.}
\newblock \emph{\bibinfo{journal}{Astronomy \& Astrophysics}} \textbf{\bibinfo{volume}{630}}, \bibinfo{pages}{A58} (\bibinfo{year}{2019}).

\bibitem{hotzel2002}
\bibinfo{author}{Hotzel, S.}, \bibinfo{author}{Harju, J.} \& \bibinfo{author}{Juvela, M.}
\newblock \emph{\bibinfo{journal}{Astronomy \& Astrophysics}} \textbf{\bibinfo{volume}{395}}, \bibinfo{pages}{L5--L8} (\bibinfo{year}{2002}).

\bibitem{lada2003}
\bibinfo{author}{Lada, C.~J.}, \bibinfo{author}{Bergin, E.~A.}, \bibinfo{author}{Alves, J.~F.} \& \bibinfo{author}{Huard, T.~L.}
\newblock \emph{\bibinfo{journal}{The Astrophysical Journal}} \textbf{\bibinfo{volume}{586}}, \bibinfo{pages}{286--295} (\bibinfo{year}{2003}).

\bibitem{Bialy2022}
\bibinfo{author}{Bialy, S.}, \bibinfo{author}{Belli, S.} \& \bibinfo{author}{Padovani, M.}
\newblock \emph{\bibinfo{journal}{Astronomy \& Astrophysics}} \textbf{\bibinfo{volume}{658}}, \bibinfo{pages}{L13} (\bibinfo{year}{2022}).

\bibitem{Nielbock2012}
\bibinfo{author}{Nielbock, M.} \emph{et~al.}
\newblock \emph{\bibinfo{journal}{Astronomy \& Astrophysics}} \textbf{\bibinfo{volume}{547}}, \bibinfo{pages}{A11} (\bibinfo{year}{2012}).

\bibitem{Roy2014}
\bibinfo{author}{Roy, A.} \emph{et~al.}
\newblock \emph{\bibinfo{journal}{Astronomy \& Astrophysics}} \textbf{\bibinfo{volume}{562}}, \bibinfo{pages}{A138} (\bibinfo{year}{2014}).

\bibitem{Lippok2016}
\bibinfo{author}{Lippok, N.} \emph{et~al.}
\newblock \emph{\bibinfo{journal}{Astronomy \& Astrophysics}} \textbf{\bibinfo{volume}{592}}, \bibinfo{pages}{A61} (\bibinfo{year}{2016}).

\bibitem{Draine1978}
\bibinfo{author}{Draine, B.~T.}
\newblock \emph{\bibinfo{journal}{The Astrophysical Journal Supplement Series}} \textbf{\bibinfo{volume}{36}}, \bibinfo{pages}{595} (\bibinfo{year}{1978}).

\bibitem{Bialy2020b}
\bibinfo{author}{Bialy, S.}  (\bibinfo{year}{2020}).
\newblock \eprint{2008.00009}.

\bibitem{Bialy2015a}
\bibinfo{author}{Bialy, S.} \& \bibinfo{author}{Sternberg, A.}
\newblock \emph{\bibinfo{journal}{Monthly Notices of the Royal Astronomical Society}} \textbf{\bibinfo{volume}{450}}, \bibinfo{pages}{4424--4445} (\bibinfo{year}{2015}).

\bibitem{Bialy2019}
\bibinfo{author}{Bialy, S.} \& \bibinfo{author}{Sternberg, A.}
\newblock \emph{\bibinfo{journal}{The Astrophysical Journal}} \textbf{\bibinfo{volume}{881}}, \bibinfo{pages}{160} (\bibinfo{year}{2019}).
\newblock \eprint{1902.06764}.

\bibitem{Sternberg2021}
\bibinfo{author}{Sternberg, A.}, \bibinfo{author}{Gurman, A.} \& \bibinfo{author}{Bialy, S.}
\newblock \emph{\bibinfo{journal}{The Astrophysical Journal}} \textbf{\bibinfo{volume}{920}}, \bibinfo{pages}{83} (\bibinfo{year}{2021}).
\newblock \eprint{2105.01681}.

\bibitem{Draine2011}
\bibinfo{author}{Draine, B.~T.}
\newblock \emph{\bibinfo{title}{Physics of the Interstellar and Intergalactic Medium}} (\bibinfo{year}{2011}).

\bibitem{Lombardi2001}
\bibinfo{author}{Lombardi, M.} \& \bibinfo{author}{Alves, J.}
\newblock \emph{\bibinfo{journal}{Astronomy \& Astrophysics}} \textbf{\bibinfo{volume}{377}}, \bibinfo{pages}{1023--1034} (\bibinfo{year}{2001}).

\bibitem{Alves2001}
\bibinfo{author}{Alves, J.~F.}, \bibinfo{author}{Lada, C.~J.} \& \bibinfo{author}{Lada, E.~A.}
\newblock \emph{\bibinfo{journal}{Nature}} \textbf{\bibinfo{volume}{409}}, \bibinfo{pages}{159--161} (\bibinfo{year}{2001}).

\bibitem{Sternberg2014}
\bibinfo{author}{Sternberg, A.}, \bibinfo{author}{Petit, F.~L.}, \bibinfo{author}{Roueff, E.} \& \bibinfo{author}{Bourlot, J.~L.}
\newblock \emph{\bibinfo{journal}{The Astrophysical Journal Supplement Series}} \textbf{\bibinfo{volume}{790}}, \bibinfo{pages}{10S} (\bibinfo{year}{2014}).
\newblock \eprint{1404.5042}.

\bibitem{Bialy2016a}
\bibinfo{author}{Bialy, S.} \& \bibinfo{author}{Sternberg, A.}
\newblock \emph{\bibinfo{journal}{The Astrophysical Journal}} \textbf{\bibinfo{volume}{822}}, \bibinfo{pages}{83} (\bibinfo{year}{2016}).

\bibitem{Sternberg1988}
\bibinfo{author}{Sternberg, A.}
\newblock \emph{\bibinfo{journal}{The Astrophysical Journal}} \textbf{\bibinfo{volume}{332}}, \bibinfo{pages}{400} (\bibinfo{year}{1988}).

\bibitem{black1987a}
\bibinfo{author}{Black, J.~H.} \& \bibinfo{author}{{van Dishoeck}, E.~F.}
\newblock \emph{\bibinfo{journal}{The Astrophysical Journal}} \textbf{\bibinfo{volume}{322}}, \bibinfo{pages}{412} (\bibinfo{year}{1987}).

\bibitem{LePetit2006}
\bibinfo{author}{Le~Petit, F.}, \bibinfo{author}{Nehme, C.}, \bibinfo{author}{Le~Bourlot, J.} \& \bibinfo{author}{Roueff, E.}
\newblock \emph{\bibinfo{journal}{The Astrophysical Journal Supplement Series}} \textbf{\bibinfo{volume}{164}}, \bibinfo{pages}{506--529} (\bibinfo{year}{2006}).

\bibitem{Padovani2022a}
\bibinfo{author}{Padovani, M.} \emph{et~al.}
\newblock \emph{\bibinfo{journal}{Astronomy \& Astrophysics}} \textbf{\bibinfo{volume}{658}}, \bibinfo{pages}{A189} (\bibinfo{year}{2022}).

\bibitem{lepetit2006a}
\bibinfo{author}{Le~Petit, F.}, \bibinfo{author}{Nehme, C.}, \bibinfo{author}{Le~Bourlot, J.} \& \bibinfo{author}{Roueff, E.}
\newblock \emph{\bibinfo{journal}{The Astrophysical Journal Supplement Series}} \textbf{\bibinfo{volume}{164}}, \bibinfo{pages}{506--529} (\bibinfo{year}{2006}).

\end{thebibliography}
\end{document}